\documentclass[useAMS]{mn2e}
\usepackage{graphicx}

\title[MONDian modelling of NGC 288] {MONDian dynamical modeling of NGC 288 with $\beta \neq 0$}

\author[X. Hernandez, R. A. M. Cort\'es and R. Scarpa] {X. Hernandez$^{1}$, R. A. M. Cort\'es$^{1}$ and R. Scarpa$^{2}$\\ 
$^{1}$Instituto de Astronom\'{\i}a, Universidad Nacional Aut\'{o}noma de M\'{e}xico,
  Apartado Postal 70--264 C.P. 04510 M\'exico D.F. M\'exico. \\
$^{2}$Instituto de Astrofsica de Canarias, C/O Via Lactea, s/n E38205—La Laguna (Tenerife), Espa\~{n}a.\\
}

\date{Released 1 June. 2016}

\pagerange{\pageref{firstpage}--\pageref{lastpage}} \pubyear{2015}

\begin{document}

\label{firstpage}

\maketitle

\begin{abstract}
NGC 288 is a diffuse Galactic globular cluster, it is remarkable in that its low density 
results in internal accelerations being below the critical MOND $a_{0}$ acceleration throughout. 
This makes it an ideal testing ground for MONDian gravity, as the details of the largely
unknown transition function between the Newtonian and modified regimes become unimportant. Further, 
exact analytical solutions exist for isothermal
spherical equilibrium structures in MOND, allowing for arbitrary values of the anisotropy parameter, $\beta$.
In this paper we use observations of the velocity dispersion profile of NGC 288, which is in fact
isothermal, as dynamical constraints on MONDian models for this cluster, where the remaining free parameters are 
adjusted to fit the observed surface brightness profile. We find the optimal fit requires $\beta =0$, 
an isotropic solution with a total mass of $3.5 \pm 1.1 \times 10^{4} M_{\odot}$.
\end{abstract}


\section{Introduction} \label{intro}
Starting with observations of $\omega$Cen in Scarpa et al. (2003), it became apparent that Galactic 
globular clusters have projected velocity dispersion radial profiles which do not fall monotonically 
with radius along Newtonian expectations for isolated systems. Rather, after an initial radial drop, 
projected velocity dispersion profiles, $\sigma(R)$, settle to constant asymptotic values. This was
then confirmed for the cases of M15 and NGC 6171 by Scarpa et al. (2004a,b), and then extended to 
NGC 7099 in Scarpa et al. (2007). Since, this generic feature has been corroborated for a growing 
sample of Galactic GCs by various independent groups, e.g. Lane et al. (2009), Lane et al. (2011).

One of the most interesting features of this asymptotic flattening in $\sigma(R)$, is that the radii 
at which they become flat, closely corresponds to those where the typical stellar acceleration 
falls below the critical MOND acceleration of $a_{0}=1.2 \times 10^{-10}m s^{-2}$ e.g. Scarpa et al. 
(2007), Hernandez \& Jimenez (2012). This last has been interpreted as evidence in favour of MOND 
scenarios (e.g. Hernandez et al. 2013), tidal disruption from the overall Galactic gravitational field 
(e.g. K\"{u}pper et al. 2010), or dynamical evolution processes internal to the Globular Clusters themselves 
(e.g.  Kennedy 2014 under Newtonian gravity).

{ The study of globular cluster
dynamics as probes of possible variations in the form of gravity and/or details of the effects of internal
dynamical evolution and tidal interactions with the Galaxy has been a topic of substantial interest over
the past few years. Dynamical modelling under MOND has been preformed by Sollima \& Nipoti (2010),
Sanders (2012) and Wu \& Kroupa 2013 under MOND, finding results in support of a MONDian interpretation,
while Lane et al. (2010) find Newtonian models yield accurate descriptions. The available data samples are
also growing, e.g. Kimmig et al (2015) and Lardo et al. (2015) preform kiematical samplings of growing sets
of globular clusters, with the recent study by Baldwin et al. (2016) giving for the first time, proper motion
kinematic profiles for a number of Galactic globular clusters.}

In support of the MONDian interpretation however, is the fact that the amplitude of 
the asymptotic $\sigma$ values closely scales with the fourth root of the total baryonic mass of the clusters 
(Hernandez et al. 2013), in accordance with MONDian predictions.  Within a Newtonian interpretation, this 
last appears as an unexplained coincidence. Further, given observed proper motions, Hernandez et al. (2013) 
also showed that Newtonian tidal radii at perigalacticon for the clusters in question, are on average a 
factor of 4 larger than those where the flattening appears.

By MONDian gravity we refer to any modified theory where at $a>a_{0}$ scales standard Newtonian 
gravity is recovered, while for $a<a_{0}$ MONDian dynamics ensue e.g. { TeVeS of Bekenstein (2004), some
of the $F(R)$ theories, Capozziello \& De Laurentis (2011), The extended Newtonian gravity of Mendoza et al.
(2011) or the covariant $F(\chi)$ of Mendoza et al. (2013).} For such theories, beyond 
a radius given by $R_{M}= (GM /a_{0})^{1/2}$, centrifugal equilibrium velocities become flat at 
$V=(GM a_{0})^{1/4}$. Similarly, for pressure supported systems, beyond around $R_{M}$, velocity dispersion
profiles will stop falling along Newtonian expectations and flatten out at a level of $\sigma_{\infty}=V/\sqrt{3}$,
with $M$ the total baryonic mass for an astrophysical system e.g. Milgrom (1984), Hernandez \& Jimenez (2012).

In more general terms, GCs are ideal testing grounds for gravity theories, as in the absence of any 
detectable gas or dust, the total baryonic mass is composed exclusively of well studied stars with 
measured metallicities and colour magnitude diagrams (CMDs). These have also been subject to detailed 
stellar population synthesis modeling tailored to each individual GC, in terms of metallicities and ages 
(e.g. McLaughlin \& van der Marel (2005).

So far,  GC MOND dynamical models have concentrated in modeling observed $\sigma(R)$ and projected 
surface brightness profiles simultaneously, under the assumption of isotropic orbits, i.e., an anisotropy 
parameter $\beta=0$, and particular forms of the MOND $\mu$ function which mediates the transition between 
the Newtonian and MOND regimes, e.g. Haghi et al. (2009), or including the effects of orbital
anisotropy for particular clusters, e.g. Sollima \& Nipoti (2010) for NGC 2419. No MONDian modelling including
anisotropy for the interesting cluster NGC 288 has been performed to date. Here, we use general anisotropic 
MOND exact analytic solutions for self-gravitating spherical stellar clusters to model simultaneously the 
projected velocity dispersion and surface brightness profiles of NGC 288.

This particular cluster is interesting, 
as its low volume density result in internal accelerations below $a_{0}$ throughout.
This last makes the details of any transition function between the Newtonian and MONDian
regimes largely unimportant, a unique case which can be studied for consistency (or otherwise) of a MONDian scenario 
through analytic dynamical models including $\beta \neq 0$. As expected under MONDian schemes, the observed velocity 
dispersion profile is flat throughout, at a value of $2.3 \pm 0.15 km s^{-1}$. This is reminiscent of the
two classes of sprial galaxies that exist, high and low surface brightness galaxies, with the low density
former ones being ``dark matter dominated'' throughout.

In section 2 we present the observations used to derive the flat velocity dispersion profile for NGC 288, 
which are then used in section 3, together with exact analytic MONDian dynamical models and the observed 
V-band surface density profile compilation for NGC 288 from Trager et al. (1995), to solve for a maximum 
likelihood dynamical model. Thus, we obtain both best fit values and confidence intervals for the mass to
light ratio, the central density of the cluster, and the $\beta$ parameter. It is interesting that the preferred
model has a total mass of $3.5 \pm 1.1 \times 10^{4} M_{\odot}$, { which implies a mass to light ratio of $1.09
\pm 0.37$, consistent with independent estimates of this ratio for the present day stellar population of this cluster
of $1.42^{+0.37}_{-0.29}$ by Kruijssen \& Mieske (2009)}. Even allowing for arbitrary values of $\beta$, the preferred
solution strongly suggests $\beta=0$. Finally, section 4 presents our conclusions.

\section{Projected velocity dispersion observations}

Initial selection of targets was based on colour, as derived
from the analysis of ESO Imaging Survey frames. A catalogue
of targets was prepared including mostly stars from the subgiant
branch down to the turn off, between 15 and 18 apparent V mag.
Observations were then obtained with FLAMES
(Pasquini et al. 2002) at the ESO VLT telescope. FLAMES is
a fibre multi-objects spectrograph, allowing the simultaneous
observation of up to 130 objects. We selected the HR9B setup
that includes the magnesium triplet covering the wavelength
range $5143 < \lambda < 5346$ at resolution R=25900. Stellar astrometry
was derived cross correlating the stellar positions on the
ESO imaging survey frames with coordinates from the US Naval Observatory 
catalogue, which proved to have the required accuracy
(0.3 arcsec) for FLAMES observations. Two different fibre
configurations were necessary to allocate all the selected stars.
For each configuration three 2700 s exposures were
obtained under good atmospheric condition (clear sky and
seeing $\sim$1 arcsec) on August 29 and 30, 2005.

Data reduction was performed within IRAF, using standard
reduction procedures. After extraction and wavelength
calibration, radial velocities were derived cross-correlating the spectra
of each target with respect to a template, the target with the
best spectrum. The two configurations shared a small number
of stars, to evaluate and eliminate possible offsets in the
velocity zero point. A posteriori, we verified that no correction
was necessary down to a level of accuracy of 250 m/s, well
below the accuracy required for our study. Finally, keeping in
mind that we are interested only on the velocity dispersion, the
global velocity zero point was derived by identifying a few lines
in the spectrum of the template. In total, 126 radial velocities
with accuracy better than $1 km/s$ were obtained. All velocities
presented here are heliocentric.

The final sample of 126 radial velocities includes virtually only
cluster members, with just 2 stars having substantially
different radial velocity from the average of the cluster.
Thus, even though the radial velocity of NGC 288 is not particularly 
high to unquestionably separate members from field stars, we expect very little contamination, 
if any at all, in our dataset. A result consistent with the high galactic latitude of this cluster.

\begin{figure}
\includegraphics[width=8.4cm,height=7.0cm]{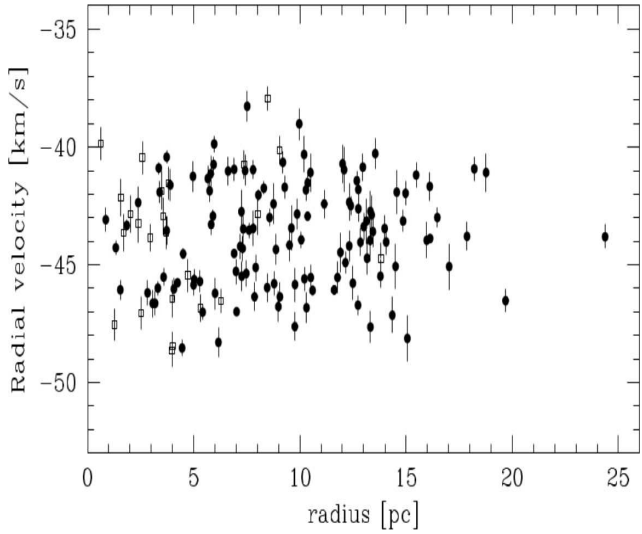}  
\caption{Radial velocity of NGC 288 members as a function of radius. Solid circles show our 124 
radial velocities, while open squares give the 24 data points from Pryor et al. (1991). The isolated 
point at r$\sim$25 pc has been excluded from the computation of the velocity dispersion.}
\end{figure}

To better constrain the velocity dispersion close to the cluster centre, we combined our data with the 
24 additional stars, mostly within 6 pc from the cluster centre and radial velocity accuracy better that 
1 km/s, from Pryor et al. (1991). After applying an offset of 2.9 km/s to match our radial velocity 
zero point, these data smoothly merge with ours in the region of overlap, showing basically the same velocity 
dispersion, as apparent in figure (1). This combined sample was used to detect evidence for ordered rotation 
in NGC 288 that might contribute to sustain the cluster. No evidence was found for ordered rotation down to the 
level of 0.5 km/s. { Any bimodality in the velocities, given the low number of bins and their errors, is not
statistically significant. Indeed, Lane et al. (2010) report ordered rotation of $0.25 \pm 0.15 km/s$, which
is consistent with no rotational support to any dynamically relevant scale.}

Velocities from our combined dataset are uniformly distributed from r$\sim$0 to r$\sim$20 pc, 
allowing us to build a well sampled velocity dispersion profile from the centre to almost 18 pc, as shown in the 
projected velocity dispersion profile given in figure (2), with error bars.
Looking at both figures (1) and (2), we see no indications of a vanishing velocity dispersion at large radii, 
rather, the dispersion is remarkably constant, being consistent with the average value of 2.3$\pm 0.15 km/s$ 
over the full range of radii covered by the data. { These results are consistent with the flat proper motion 
dispersion profile by Baldwin et al. (2016), who report a dispersion of $2.7 \pm 0.4 km/s$ in the central 4 pc of
the cluster. The velocity dispersion profile of Lane et al. (2010) is quantitatively consistent with our data
internal to 12 pc, the slight fall reported by those authors at larger radii (two points) might be due to
contamination effects towards the low surface brightness outer regions.}

Figure (2) also shows the best fit constant velocity, straight line. For a Newtonian comparison we also give a Plummer
model fit having the observed half-light radius of NGC 288, and leaving $\sigma_{0}$ as a free parameter, thus, a one parameter
fit as in the isothermal case. Dejonghe (1987) shows that the projected velocity dispersion profile of a Plummer model satisfies
$\sigma(R)^{2}=\sigma_{0}^{2}[1+(R/R_{h})^{2}]^{-1/2} $, where $R_{h}$ is the observed half light radius and $\sigma_{0}$ the central
velocity dispersion. The optimal fit in this last case appears systematically towards the upper edges of the observed velocity
dispersion confidence intervals in the central regions of the cluster, and towards the lower edges of these confidence
intervals in the outer zones. { Not surprisingly, the isothermal fit is slightly better than the Newtonian one;
optimal $\chi^{2}$ values increase by $10\%$ from the isothermal case of $\chi^{2}=0.50$ to $\chi^{2}=0.55$ for
the Plummer model}. This, and the already mentioned lack of relevance of any unknown transition function, motivate a MONDian study
of this cluster.

\begin{figure}
\includegraphics[width=8.4cm,height=7.0cm]{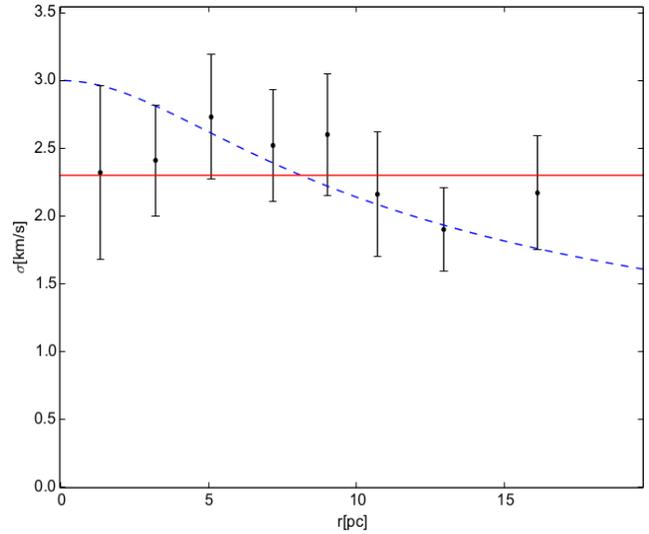}  
\caption{The radial velocity dispersion profile as derived from our 124 radial velocities together with 
24 velocities from Pryor et al. (1991). The abscissa of each point is the average of the points in the bin. 
Error bars give the 1$\sigma$ uncertainty on the dispersion. The central velocity dispersion is 
from Pryor \& Meylan (1993). The optimal isothermal fit is given by the horizontal line, while the curve shows the optimal
Newtonian Plummer fit for the observed half-light radius of NGC 288, having a higher $\chi^{2}$ than the MONDian
isothermal model.}
\end{figure}

\section{MONDian isothermal model for NGC 288}

Following the solution proposed by Milgrom (1984), we can solve the hydrostatic equilibrium equation 
for a polytropic equation of state $P=K\rho^\gamma$, which is:

\begin{equation} \label{hidro}
K\gamma\rho^{\gamma-2}\frac{d\rho}{dr} = -\nabla \phi. 
\end{equation}

\noindent If we consider the case of local isothermal conditions, then $\gamma=1$ and $K=\sigma^2$. Then, by 
putting equation \ref{hidro} in terms of mass, given $4 \pi r^{2} \rho = dM(r)/dr$, turns it into:

\begin{equation} \label{hidro2}
\sigma^2\left[
\left(\frac{dM(r)}{dr}\right)^{-1}\frac{d^2M(r)}{dr^2}-\frac{2}{r}
\right] = -\nabla\phi.
\end{equation}

\noindent However, when anisotropy is considered in the model, in terms of $0 \leq \beta \leq 1$, 
$\beta \equiv 1 - \sigma_t/\sigma_r$, being $\sigma_t$ the tangential and $\sigma_r$ the radial components 
of velocity dispersion, eq. \ref{hidro} becomes:

\begin{equation} \label{hidro3}
\sigma_r^2 \frac{d\rho}{dr} + \frac{2\rho\beta\sigma_r^2}{r} = -\rho\nabla\phi,
\end{equation}

\noindent which in terms of mass reads:
\begin{equation} \label{hidro4}
\sigma_r^2\left[
\left(\frac{dM(r)}{dr}\right)^{-1}\frac{d^2M(r)}{dr^2}-\frac{2(1-\beta)}{r}
\right] = -\nabla\phi.
\end{equation}

\begin{figure}
\includegraphics[width=8.4cm,height=7.0cm]{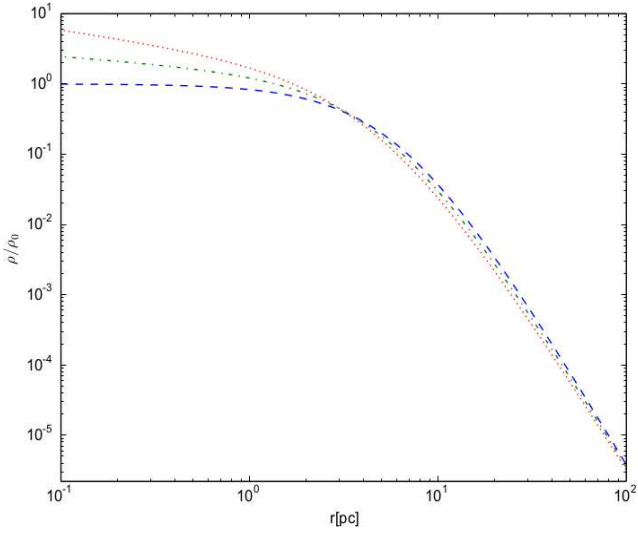}  
\caption{Theoretical dimensionless density profiles for $\rho_0=100M\odot$, $\sigma_r=2.3\mathrm{km/s}$ and
$\beta=\{0.0,0.05,0.1\}$, bottom to top on the y-axis, respectively.
Notice that for $\beta=0$ the density is practically constant within 1pc, but for other 
$\beta$ values, density increases at smaller radii.}
\end{figure}

\noindent Until now, the gravitational potential has not been explicitly specified. If we use a MONDian 
force law, assuming the cluster always remains in the low acceleration regime, as is the case for NGC 288, we obtain,

\begin{equation} \label{hidromond}
\sigma_r^2\left[
\left(\frac{dM(r)}{dr}\right)^{-1}\frac{d^2M(r)}{dr^2}-\frac{2(1-\beta)}{r}
\right] = -\frac{[GM(r)a_0]^{1/2}}{r},
\end{equation}

\noindent where $a_0$ is the MOND critical acceleration. If we define $M_0 \equiv 9\sigma_r^4/G a_0$ 
and $r_0 \equiv (G M_0 / a_0)^{1/2}$ for simplicity, we can treat eq.(5) using the dimensionless variables 
${\cal M} \equiv M/M_0$ and $R \equiv r/r_0$ to get the dimensionless equation:

\begin{equation}
R \frac{d^{2}{\cal M}}{d R^{2}} = \frac{d {\cal M}}{dR} (2-2\beta-3{\cal M}^{1/2}),
\end{equation}

\noindent with solution,

\begin{equation}
{\cal M}=\left( \frac{2\beta -3} {2+ {\cal R}^{(2\beta-3)/2}} \right)^{2}.
\end{equation}

\noindent (Milgrom 1984), where ${\cal R} \equiv R/R_S$, and $R_S$ is a scale radius at which, for the perfectly 
isotropic case ($\beta=0$), ${\cal M}=1$. Notice a finite total mass of  ${\cal M}= (3-2\beta)^{2}/4$ results.
In the above, the boundary condition ${\cal M}(0)=0$ has been assumed.

Differentiation of eq. (7) yields the exact volumetric density profile, in terms of the central density $\rho_0$:
\begin{equation} \label{voldens}
\rho({\cal R}) = \frac{\rho_0}{[2{\cal R}^{(3-2\beta)/2}+1]^3{\cal R}^{2\beta}}
\end{equation}

\begin{figure}
\includegraphics[width=8.4cm,height=7.0cm]{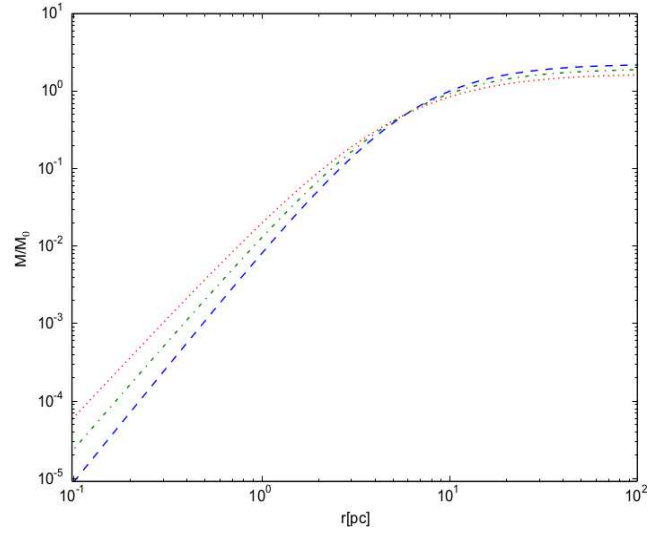}  
\caption{Theoretical dimensionless mass profiles for $\rho_0=100M\odot$, $\sigma_r=2.3\mathrm{km/s}$ and
$\beta=\{0.0,0.05,0.1\}$, bottom to top on the y-axis, respectively. Mass converges to $(3-2\beta)^2M_{0}/4$.}
\end{figure}

\noindent Notice that ${\cal R}$ is not a physical radius, in general $R_S = (3-2\beta)(4\pi r_0^3 \rho_0/M_0)^{-1/3}$,
which recovers an expression in terms of physical distances.

In figures (3) and (4) we give examples of the mass profiles which result from equations (7) and (8), showing $\rho(r)/\rho_{0}$
and $M(r)/M_{0}$ in logarithmic plots, for values of $\beta= 0.0, 0.05$ and $0.1$, bottom to top in the y-axis of the two plots,
respectively. We see $\beta=0$ yielding a constant density central core out to a few pc, for the value of $\sigma=2.3 km s^{-1}$
adopted here, followed by an approximately isothermal region, which then steepens to produce a convergent total mass. As $\beta$
is increased, a central density spike evolves, while the outer structure remains largely unchanged.

Before comparing with the observed surface brightness profile of NGC 288, we must not only project the analytical volumetric
density profile and assume a M/L ratio, but also, be aware that the continuous matter distribution assumption 
inherent to the model presented above , will break down internal to a critical radius. Given typical stellar masses of
$1 M_{\odot}$, at radii internal to which the volumetric mass of the model falls below $1 M_{\odot}$, we will either find
a star or not find a star, rather than find a continuous distribution of mass totalling say, 0.8 stars. Even before such
radii, the continuous distribution assumption will begin to break down. Since such assumption requires a large (and constant)
number of stars within any given radius, we can estimate the inner validity critical radius as the one internal to which
the total model mass falls below $2^{3} M_{\odot} = 8 M_{\odot}$, outside of which fluctuations will be minor and the continuous 
matter distribution assumption will hold. { In what follows we refer to the above critical radius as $r_{c}$.}

Now, having the volumetric density profile, in order to get the corresponding surface density profile as projected in the sky, 
we use:
\begin{equation} \label{surfdens}
\Sigma(R_{p}) = 2 \int_0^{\pi/2}\rho\left(\frac{R_{p}}{\cos^2\theta}\right) \frac{R_{p}}{\cos^2\theta} d\theta
\end{equation}

where $R_{p}$ denotes the projected distance from the centre of the object. By substituting eq.(8) into 
eq.(9), we obtain:

\begin{equation}
\Sigma(R_{p}) = \frac{2R_{p}\rho_0}{{\cal R}^{2\beta}}\int_0^{\pi/2}\frac{(\cos\theta)^{2(\beta-1)}}{[2(\frac{{\cal R}}{\cos\theta})
^{(3-2\beta)/2}+1]^3} d\theta
\end{equation}

\begin{figure}
\includegraphics[width=8.4cm,height=7.0cm]{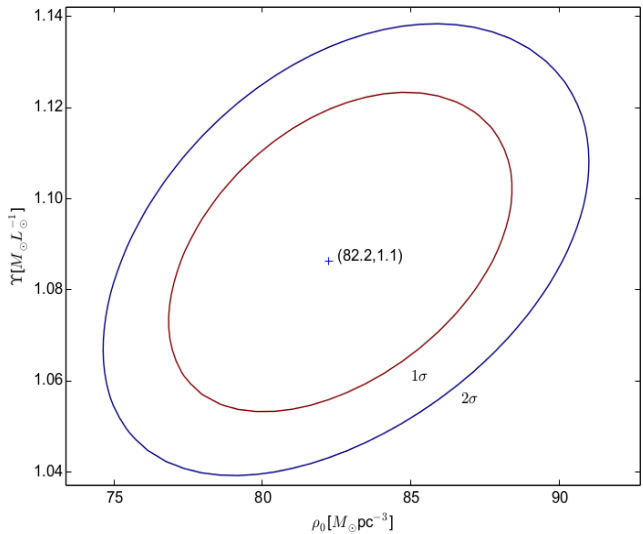}  
\caption{Maximum likelihood value for the $\beta=0.0$ model, and $1\sigma$ and $2\sigma$ contours.}
\end{figure}

In this expression, ${\cal R}(R)=R/(r_0 R_S)$. Now, the integral in this previous equation can be evaluated numerically, 
and its value depends only on ${\cal R}={\cal R}(R,\rho_0,\sigma_r)$.

Then, we can translate this surface density profile into a surface brightness profile, always considering a fixed 
mass to light ratio $\Upsilon$. Then the surface brightness will be:

\begin{equation} \label{surbright}
\mu\left[\frac{\mathrm{mag}}{\mathrm{arcsec}^2}\right] = M_{i\odot} + 21.572 - 2.5\log(\Sigma/\Upsilon)
\end{equation}

\noindent where $M_{i\odot}$ is the solar absolute magnitude in a given band and $\Sigma$ is in units of
[$\mathrm{L}_\odot/\mathrm{pc}^2$]. This way, $\mu$ is completely determined by the parameters $\sigma_r$,$\rho_0$,$\Upsilon$
and $\beta$.

Since NGC 288 has a dispersion profile consistent with a constant value of $\sigma_r = 2.3$ km/s, 
it is a good candidate for an isothermal cluster that can be modelled via the exact MONDian self-gravitating profile
shown above. Fitting the results of eq. (11) to the { V-band} surface brightness data compiled in Trager et al. (1995) 
through a maximum likelihood method, we can retrieve the most likely parameters for the cluster under a MONDian isothermal
model. Having fixed the dispersion velocity, only three free parameters remain, ($\rho_0$,$\Upsilon$ and $\beta$). Then,
by successively choosing fixed values of $\beta=\{0.0, 0.05, 0.1, 0.15, 0.2, 0.3, 0.4\}$, we explored the $(\rho_0 , \Upsilon)$
parameter space to obtain the best match. The maximum  likelihood map is shown in figures (5) and (6) for $\beta=0.0$ and
$\beta=0.05$, respectively, with isocontours for $1\sigma$ and $2\sigma$ confidence intervals and optimal values shown
in the centre.

%
%
%

Figure (5) shows the results of the maximum likelihood analysis in the $(\rho_{0}, M/L)$ parameter space, for the case
of $\beta=0.0$. We see an optimal solution at $\rho_{0}=82.2 M_{\odot} pc ^{-3}$ and an optimal $M/L=1.09$. { All $M/L$
ratios mentioned refer to V-band values.} The figure also gives 1 and 2 $\sigma$ confidence intervals, approximately
$\pm 8 M_{\odot} pc ^{-3}$ and $\pm 0.05$ for the 1$\sigma$ case in $\rho_{0} $ and $M/L$, respectively. 
Figure (6) is equivalent to figure (5), but this time for the
case of $\beta=0.05$. We see that although the inferred $M/L$ values are consistent with what results assuming $\beta=0$,
in this case somewhat smaller $\rho_{0}$ values result. As is also evident from figure (5), a slight positive correlation
appears between the values of the two inferred quantities; if one chooses $\rho_{0}$ values towards the high end of the
confidence interval, $M/L$ values towards the high end of their corresponding confidence interval are required.

Cuts where made at other values of $\beta \neq 0$, resulting in qualitatively similar plots. It is interesting that the
actual maximum log(likelihood) values obtained at all $\beta \neq 0$ cuts are always more than 0.5 units below the maximum
obtained in the $\beta =0 $ case. Thus, an isotropic solution is preferred by the full maximum likelihood fit to the
observed surface brightness profile, with a very small uncertainty smaller than 0.05. Even allowing for arbitrary
values of the anisotropy parameter, the dynamical models used best represent the structure of the observed cluster for
an isotropic solution.

\begin{figure}
\includegraphics[width=8.4cm,height=7.0cm]{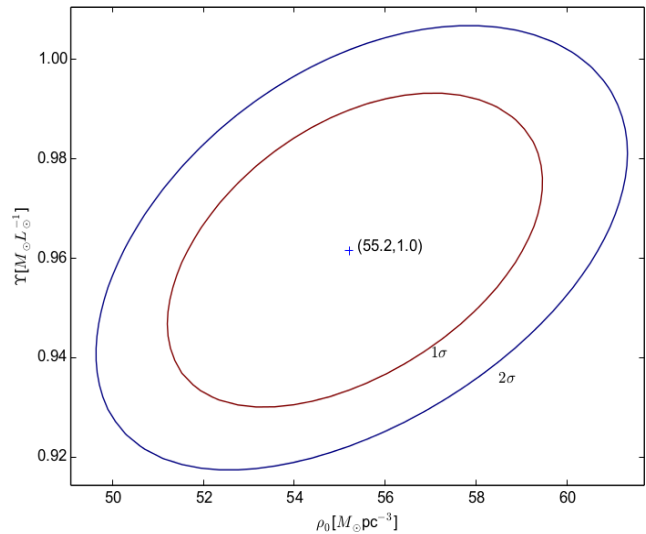}  
\caption{Maximum likelihood value for the $\beta=0.05$ model, and $1\sigma$ and $2\sigma$ contours.}
\end{figure}

The whole exploration of the 3-dimensional parameter space of $(\rho_{0}, M/L, \beta)$ described above was repeated, 
changing the input value of the stellar velocity dispersion, within the inferred confidence interval of this
observed quantity of $2.3 \pm 0.15 km/s$. Thus, the full 4-dimensional parameter space was sampled. Results where 
qualitatively equivalent to what was described above, again, resulting in significantly lower maximum likelihood 
values for $\beta \neq 0$, and interestingly, lower overall maximum likelihood values even at $\beta=0$, when
the stellar velocity dispersion was shifted away from its central value. Thus, the isotropic solution is robust
to uncertainties in the observed stellar velocity dispersion, and interestingly, the observed best fit stellar
velocity dispersion value is optimal also in terms of the overall structure of the globular cluster, when 
modelled through the dynamical models used here. By noting the very flat central surface density profile of
NGC 288 we can understand the maximum likelihood rejection of $\beta \neq 0$ models, as they all lead to a central
cusp in the density profiles.

{ In general, the intrinsic orbital structure of a stellar system is expected to be isotropic, either as a result
of retaining the kinematics of an original gas phase, or due to a process of violent relaxation. If a substantial
infall of stars then occurs, distribution functions with a degree of radial bias ($\beta>0$) are expected, e.g.
Binney \& Tremaine (1987). Tangentially biased stellar distributions, $\beta<0$, are much harder to envision. Indeed,
the only mechanism expected to yield such orbital structure is core scouring, where the presence of a super massive
black hole binary in the centre of an elliptical galaxy ejects stars with orbits having close approaches to the
central region, leaving the black hole's inner region of dynamical influence devoid of stars on preferentially radial
orbits, and hence having a tangentially biased ($\beta<0$) distribution function, e.g. as inferred through N-body
simulations by Milosavljevic \& Merritt (2001) and recently observed by Thomas et al. (2014). A similar effect appears
when a central black hole is allowed to grow adiabatically at the centre of an initially isotropic stellar population,
e.g. Goodman \& Binney (1984), Quinlan et al. (1995). 

Our results would suggest that stars in NGC 288 have retained the kinematics of an initial gaseous phase for the system,
and hence the best fit $\beta=0$ result obtained. Some improvement in the central regions would result from taking
$\beta<0$; we do not consider that option as core scouring by a central black hole (indeed the very existence of a
central black hole of any importance) appears unlikely in the very low density NGC 288 studied here, and further,
such improvement in the resulting surface density profile would be mostly confined to the innermost region where in
any case the continuous mass distribution assumption breaks down, $R<r_{c}$.}

Finally, in figure (7) we show the resulting surface density profile for the best fit model, continuous line.
The variations due to the $1\sigma$ confidence intervals lie within the thickness of the line. The dots with error bars
show the collection of observed surface brightness measurements from Trager et al. (1995) for NGC 288, which are clearly
well represented by the best fit parameters and the MONDian dynamical model used. For comparison,
the dashed line gives the best fit model for $\beta=0.3$, which is appreciably less of a good fit,
the slight offset and the many observed points available allow its rejection at a $1\sigma$ level.

\begin{figure}
\includegraphics[width=8.4cm,height=7.0cm]{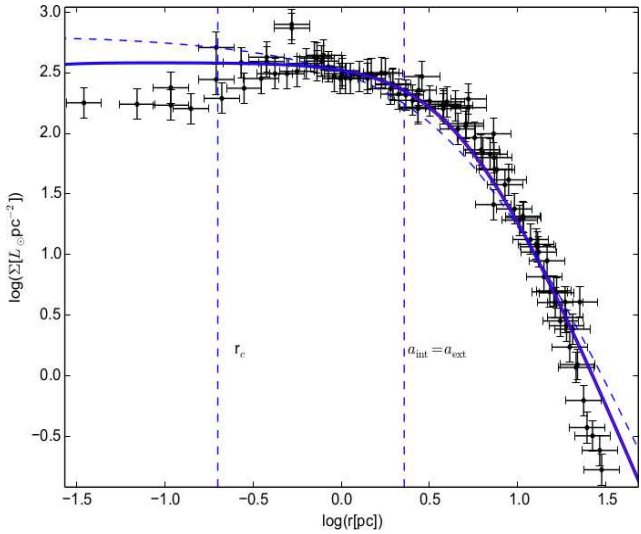}  
\caption{Maximum likelihood surface brightness model for $\beta=0.0$, $1\sigma$ variations fall within the thickness
of the solid line, and the observational data. The dashed line represents the $\beta=0.3$ model, which evidently falls 
out of the $1\sigma$ band. The vertical $r_{c}$ dashed line represents the critical radius that contains less than 
$8M_\odot$, a limit internal to which the continuous density distribution approximation begins to break down. 
The $a_{int}=a_{ext}$ vertical line gives the point exterior to which the external Galactic acceleration field dominates 
over the cluster's internal acceleration, we see no evident change in the goodness of fit on crossing this line.}
\end{figure}

We have included two dashed vertical lines indicating critical radii for the problem, the first, $r_{c}$, indicates
the critical radius internal to which the best fit model contains only $8  M_{\odot}$, as mentioned earlier, an
estimate of the region interior to which the continuous density distribution assumption will clearly break down. 
It is interesting to note that it is interior to precisely this critical radius that the surface brightness 
observations begin to noticeably diverge from the model profiles. Thus, the appearance of an inner central dip in the
observed surface brightness profile, probably corresponds to the inner region where statistical fluctuations in the 
sampling of the discrete stellar distribution begin to become important.

{ NGC288 lies at a galactocentric radius of 11 kpc. Assuming a flat rotation curve value for the Milky
Way at that radius of 200 km/s, results in an external acceleration of $1.18 \times 10^{-10} m/s^2$,
essentially $a_{0}$. Thus, this cluster does not lie in the deep MOND regime. It is a matter
of debate whether MOND as such is a final, definitive alternative to GR, or just an intriguing set of
empirical results yielding clues into the deeper nature of gravity. The predicted external field effect of MOND
as such has never been unambiguously observed, and indeed it is not a necessity under all MONDian theories (e.g.
the extended Newtonian gravity of Mendoza et al. 2011 or MONDian $F(R)$ covariant theories, e.g.
Capozziello \& De Laurentis 2011).}

A second critical radius appears at $a_{int}=a_{ext}$, the point where the square of the velocity dispersion divided 
by the radius is equal to $\sigma_{H}^{2}/R_{orb}$, where $\sigma_{H}$ and $R_{orb}$ are the $160 km/s$ and $11 kpc$
of the equilibrium velocity dispersion of tracers in the Galactic halo, and the Galactic orbital radius of NGC 288, 
respectively. Thus, the second vertical dashed line gives the radius external to which the external acceleration
field of the Galaxy dominates over the internal acceleration of NGC 288 itself. Within a strict MOND model,
a regime change is expected whenever the internal acceleration of a system changes from being larger to smaller than
the external acceleration field. It is reassuring of the modeling performed here that although no such external
field effect has been considered, the adequacy of the fit to the observed surface brightness observations does not
present any change or distinctive feature on crossing the $a_{int}=a_{ext}$ radius.

Finally, although the maximum likelihood fitting results in only a $4.5 \%$ error on the inferred mass to light radius,
given the scaling of the model mass with the fourth power of the velocity dispersion, the $0.15 km/s$ confidence
interval on the velocity dispersion of the cluster results in a mass uncertainty of $29 \%$, making the total
$1\sigma$ error on the inferred mass of the cluster, or the inferred mass to light ratio, of $33.5 \%$.
{ Thus, the optimal total mass recovered is of $3.5 \pm 1.1 \times 10^{4} M_{\odot}$, not surprisingly, slightly smaller than 
the $4.89^{+0.16}_{-0.2} \times 10^{4} M_{\odot}$ of Newtonian dynamical models for this cluster, e.g. McLaughlin \& van der
Marel (2005) or the $4.4^{+0.32}_{-0.26} \times 10^{4} M_{\odot}$ of Lane et al. (2010). Our resulting stellar $M/L$ of
$1.09 \pm 0.37$ falls below the expected value for a single stellar population having the inferred age and metallicity of
NGC 288, $M/L_{SSP} =2.2 \pm 0.08$, e.g. Kruijssen \& Mieske (2009), but is consistent to $1\sigma$ with the value predicted
for this cluster of $1.42^{+0.37}_{-0.29}$ by the same authors, after accounting for various dynamical effects such as the
selective ejection of low mass stars and the preferential loss of stellar remnants. Indeed, it is expected for present day
$M/L$ ratios in globular clusters to fall slightly below single stellar population estimates, due to the dynamical effects
mentioned above (Kruijssen \& Lamers 2008), which surely also apply under MONDian scenarios, albeit with differences in the
detail which remain to be estimated.}

\section{Final remarks}

We have constructed dynamical MONDian models taking as constraints both the observed velocity dispersion
and surface brightness profiles for NGC 288, allowing for arbitrary values of the anisotropy parameter $\beta$.
The optimal solution favours $\beta=0$, together with a final surface brightness profile closely matching the
observed one. No transition is evident when comparing with the data on crossing the point where the internal
acceleration of the cluster drops below the external acceleration of the Galaxy.



\section*{acknowledgements}

The authors acknowledge the constructive input from an anonymous referee as important towards
reaching a more clear and complete final version. This work was supported in part by DGAPA-UNAM
PAPIIT IN-100814 and CONACyT.

\end{document}